\documentstyle[]{article}
\begin{document}
\title{Quantum probabilities for time-extended alternatives }
\author{Charis Anastopoulos\footnote{anastop@physics.upatras.gr}\\
{\small Department of Physics, University of Patras, 26500 Patras,
Greece}
\\
 and \\ Ntina Savvidou \footnote{ntina@imperial.ac.uk} \\
 {\small  Theoretical Physics Group, Imperial College, SW7 2BZ,
London, UK}}

\maketitle

\begin{abstract}
We study the probability assignment for the outcomes of
time-extended measurements. We construct the class-operator that
incorporates the information about a generic time-smeared quantity.
These class-operators are employed for the construction of
Positive-Operator-Valued-Measures for the time-averaged quantities.
The scheme highlights the distinction between velocity and momentum
in quantum theory. Propositions about velocity and momentum are
represented by different class-operators, hence they define
different probability measures. We provide some examples, we study
the classical limit and we construct probabilities for generalized
time-extended phase space variables.

\end{abstract}
\section{Introduction}

In this article we  study the probability assignment for quantum
measurements of observables that take place in finite time. Usually
measurements are treated as instantaneous. One assumes that the
duration of interaction between the measured system and the
macroscopic measuring device is much smaller than any macroscopic
time scale characterising the behaviour of the measurement device.
Although this is a reasonable assumption, measurements that take
place in a macroscopically distinguishable time interval are
theoretically conceivable, too. In the latter case one expects that
the corresponding probabilities would be substantially different
from the ones predicted by the instantaneous approximation.
Moreover, the consideration of the duration of the measurement as a
determining parameter allows one to consider observables whose
definition explicitly involves a finite time interval. Such
observables may not have a natural counterpart when restricted to
single-time alternatives. In what follows, we also study  physical
quantities whose definition involves time-derivatives of single-time
observables.

There are different procedures we can follow for the study of
finite-time measurements. For example, one may employ standard
models of quantum measurement and refrain from taking the limit of
almost instantaneous  interaction between the measuring system and
the apparatus \cite{PeWo85}.  However, there is an obvious drawback.
For example, a measurement of  momentum can be implemented by
different models for the measuring device. They all give essentially
a probability that is expressed in terms of momentum spectral
projectors (more generally positive operators). However, if one
considers a measurement of finite duration, it is not obvious to
identify the physical quantity of the measured system to which the
resulting probability measure corresponds.

This problem is especially pronounced when one considers
measurements of relatively large duration. For the reason above, we
choose a different starting point: we identify time-extended
classical quantities and the we construct corresponding operators
that act on the Hilbert space of the measured system. A special case
of such observables are  quantities that are smeared in time. If an
operator $\hat{A}$ has (generalised) eigenvalues $a$, then we
identify a probability density for its time-smeared values $\langle
a \rangle_f = \int_0^T dt \, a_t f(t)$. Here $f(t)$ is a positive
function defined on the interval $[0, T]$. The special case $f(t) =
\frac{1}{T}$ corresponds to the usual notion of time-averaging.

Having identified the operators that represent the time-extended
quantities, it is easy to construct the corresponding probability
measure for such observables using for example, simple models for
quantum measurement.

Our analysis is facilitated by a comparison with the decoherent
histories approach to quantum mechanics \cite{Gri84, Omn8894,
GeHa9093, Har93a}. The identification of operators that correspond
to time-extended observables is structurally similar to the
description of temporally extended alternatives in the decoherent
histories approach  \cite{Har, scc, IL, Sav99, MiHa96, Ha02,
BoHa05}. The physical context is different, in the sense that the
decoherent histories scheme attempts the description of individual
closed systems, while the study of measurements we undertake here
involves---by necessity---the consideration of open systems.
However,
 the mathematical descriptions are very closely related.

A history is defined as a sequence of propositions about the
physical system at successive moments of time. A proposition in
quantum mechanics is represented by a projection operator; hence, a
general $n$-time history $\alpha$ corresponds to a string of
projectors $\{\hat{P}_{t_1}, \hat{P}_{t_2}, \ldots, \hat{P}_{t_n}
\}$. To determine the probabilities associated to these histories we
define the class operator $\hat{C}_{\alpha}$,
\begin{equation}
\hat{C}_{\alpha} =\hat{U}^{\dagger}(t_1) \hat{P}_{t_1} \hat{U}(t_1)
\ldots \hat{U}^{\dagger}(t_n) \hat{P}_{t_n} \hat{U}(t_n),
\label{ccll}
\end{equation}
where $\hat{U}(t) = e^{-i \hat{H}t}$ is the evolution operator for
the system. For a pair of histories $\alpha$ and $\alpha'$, we
define the decoherence functional
\begin{equation}
d(\alpha, \alpha') = Tr \left( \hat{C}_{\alpha}^{\dagger}
\hat{\rho}_0 \hat{C}_{\alpha'} \right). \label{decfun}
\end{equation}
A key feature of the decoherent histories scheme is that
probabilities can be assigned to an exclusive and exhaustive set of
histories only if the decoherence condition
\begin{eqnarray}
d(\alpha, \alpha') = 0, \; \alpha \neq \alpha'
\end{eqnarray}
holds. In this case one may define a probability measure on this
space of histories
\begin{eqnarray}
p(\alpha) = Tr \left( \hat{C}_{\alpha}^{\dagger} \hat{\rho}
\hat{C}_{\alpha} \right). \label{pmeas}
\end{eqnarray}
One of the most important features of the decoherent histories
approach is its rich logical structure: logical operations between
histories can be represented in terms of algebraic relations between
the operators that represent a history. This logical structure is
clearly manifested in the History Projection Operator (HPO)
formulation of decoherent histories \cite{I94}. In this paper we
will make use of the following property. If $\{\alpha_i\}$ is a
collection of mutually exclusive histories, each represented by the
class operator $\hat{C}_{\alpha_i}$ then the coarse-grained history
that corresponds to the statement that any one of the histories $i$
has been realised is represented by the class operator $\sum_i
\hat{C}_{\alpha_i}$. This property has been employed by Bosse and
Hartle \cite{BoHa05},  who define class operators corresponding to
time-averaged position alternatives using path-integrals. A similar
construction in a slightly different context is given by Sokolovski
et al \cite{So98, LS00}---see also Ref. \cite{Caves}.

Our first step  is to generalise  the results of \cite{BoHa05} by
constructing such class operators for the case of a generic
self-adjoint operator $\hat{A}$ that are smeared with an arbitrary
function $f(t)$ within a time interval $[0, T]$. This we undertake
in section 2.

In section 3, we describe a toy model for a time-extended
measurement. It leads to a probability density for the measured
observable that is expressed solely in terms of the class operators
$\hat{C}_{\alpha}$. The same result can be obtained without the use
of models for the measurement device through a purely mathematical
argument. We identify generic Positive-Operator-Valued Measure
(POVM) that is bilinear with respect to the class operators
$\hat{C}_{\alpha}$ and compatible with Eq. (\ref{pmeas}).

The result above  implies that $\hat{C}_{\alpha}$ can be employed in
two different roles: first, as ingredients of the decoherence
functional in the decoherent histories approach and second, as
building block of a POVM in an operational approach to quantum
theory. The same mathematical object plays two different roles: in
\cite{BoHa05} class operators corresponding to time-average
observables are constructed for use within the decoherent histories
approach, while the same objects are used in \cite{LS00} for the
determination of probabilities of time-extended position
measurements.

The approach we follow allows the definition of more general
observables. Within the context of the HPO approach, velocity and
momentum are represented by different (non-commuting) operators:
they are in principle distinguishable concepts \cite{Sav99}.

 In
section 4, we show that indeed one may assign  class operators to
alternatives corresponding to values of velocity that are distinct
from those corresponding to values of momentum. These operators
coincide at the limit of large coarse-graining (which often
coincides with the classical limit). In effect, two quantities that
coincide in classical physics are represented by different objects
quantum mechanically. It is quite interesting that the POVMs
corresponding to velocity are substantially different from those
corresponding to momentum. At the formal level, it seems that
quantum theory allows  the existence of instruments which are able
to distinguish between the velocity and momentum of a quantum
particle. {\em A priori}, this is not surprising: in single-time
measurements, velocity cannot  be defined as an independent
variable. For extended-in-time measurements, it is not inconceivable
that one type of detector responds to the rate of change of the
position variable and another to  the particle's momentum. Whether
this result is a mere mathematical curiosity, or whether one can
design experiments that will demonstrate this difference completely
will be addressed in a future publication. In section 4 we also
study more general time-extended measurements, namely ones that
correspond to time-extended phase space properties of the quantum
system.

\section{Operators representing time-averaged quantities}

\subsection{The general form of the class operators}

We construct the class operators that correspond to the proposition
``the value of the observable $\hat{A}$, smeared with a function
$f(t)$ within a time interval $[0, T]$, takes values in the subset
$U$ of the real line ${\bf R}$.''

We denote by $a_t$ a possible   value of the observable $\hat{A}$ at
time $t$. Then at the continuous-time limit the time-smeared value
$A_f$ of $\hat{A}$ reads $A_f :=  \int_0^T a_t f(t) dt$. Note that
for the special choice $f(t) = \frac{1}{T} \chi_{[0, T]}(t)$, where
$\chi_{[0, T]}$ is the characteristic function of the interval $[0,
T]$, we obtain the usual notion of the time-averaged value of a
physical quantity.

There are two benefits from the introduction of a general function
$f(t)$.  First,  it can be chosen to be a continuous function of
$t$, thus allowing the consideration of more general `observables';
for example observables that involve the time derivatives of $a_t$.
Second, when we consider measurements, the form of $f(t)$ may be
determined by the operations we effect on the quantum system. For
example, $f(t)$ may correspond to the shape of an electromagnetic
pulse acting upon a charged particle during measurement.

To this end, we construct the relevant class operators in a
discretised form. We partition the interval $[0, T]$ into $n$
equidistant time-steps $ t_1, t_2, \ldots, t_n $. The integral
$\int_0^T dt f(t) a_t$ is obtained as the continuous limit of
$\delta t \sum_i f(t_i) a_{t_i} = \frac{T}{n} \sum_i f(t_i)
a_{t_i}$.

For simplicity of exposition we assume that the operator $\hat{A}$
has discrete spectrum, with eigenvectors $|a_i\rangle$ and
corresponding eigenvalues $a_i$ \footnote{The generalization of our
results for continuous spectrum is straightforward.}. We write
$\hat{P}_{a_i} = | a_i \rangle \langle a_i|$. By virtue of Eq.
(\ref{ccll}) we construct the class operator
\begin{eqnarray}
\hat{C}_{\alpha} =  e^{i\hat{H}T/n} |a_1 \rangle \langle
a_1|e^{i\hat{H}T/n} |a_2 \rangle \langle a_2 | \ldots \langle
a_{n-1}|e^{i \hat{H}T/n}|a_n \rangle \langle a_n| \label{ca}
\end{eqnarray}
that represents the history $\alpha = ( a_1, \ldots , a_n)$.

 The
proposition  ``the time-averaged value of $\hat{A}$ lies in a subset
$U$ of the real line'' can be expressed by summing over all
operators of the form of Eq. (\ref{ca}), for which $\frac{T}{n}
\sum_i f(t_i) a_i \in U $,

\begin{eqnarray}
\hat{C}_U = \sum_{a_1, a_2, \ldots, a_n}  \chi_U\left(\frac{T}{n}
\sum_i f(t_i) a_i \right) \hspace{3cm} \nonumber \\
\times\, e^{i\hat{H}T/n}|a_1 \rangle \langle a_1|e^{i\hat{H}T/n}
|a_2 \rangle \langle a_2 | \ldots \langle a_{n-1}|e^{i
\hat{H}T/n}|a_n \rangle \langle a_n|. \label{class}
\end{eqnarray}

If we partition the real axis of values of the time-averaged
quantity $A_f$ into mutually exclusive and exhaustive subsets $U_i$,
the corresponding alternatives for the value of $A_f$ will also be
mutually exclusive and exhaustive.

Next, we  insert  the Fourier transform $\tilde{\chi}_U$ of $\chi_U$
defined by
\begin{eqnarray}
\chi_{U}(x) := \int \frac{dk}{2 \pi} e^{ikx} \tilde{\chi}_U(k)
\end{eqnarray}
 into
Eq. (\ref{class}). We thus obtain
\begin{eqnarray}
\hat{C}_U =  \int \frac{dk}{2 \pi} \tilde{\chi}_U(k) e^{-i
\hat{H}T/n} \left( \sum_{a_1} e^{-ik T f(t_1) a_1/n} |a_1 \rangle
\langle a_1| \right) e^{i\hat{H}T/n} \ldots \nonumber \\ \times e^{i
\hat{H}T/n} \left( \sum_{a_n} e^{-ik T f(t_n) a_n/n} |a_n \rangle
\langle a_n| \right).
\end{eqnarray}

By virtue of the spectral theorem we have
\begin{eqnarray}
\sum_{a_i} e^{ik T f(t_i) a_i/n} |a_i \rangle \langle a_i | = e^{i k
f(t_i) \hat{A}/n}.
\end{eqnarray}
Hence,
\begin{eqnarray}
\hat{C}_U =  \int \frac{dk}{2 \pi}\,  \tilde{\chi}_U(k)
\prod_{i=1}^n [
 e^{i\hat{H} T/n}e^{ik f(t_i) \hat{A}T/n} ]. \label{class4}
\end{eqnarray}

From Eq. (\ref{class4}) we obtain
\begin{eqnarray}
\hat{C}_U = \int_U da \; \hat{C}(a), \label{class5}
\end{eqnarray}
where
\begin{eqnarray}
\hat{C}(a) := \int \frac{dk}{2 \pi} e^{-i ka} \hat{U}_f(T, k),
\label{class2}
\end{eqnarray}
and where
\begin{eqnarray}
\hat{U}_f(T,k) := \lim_{n \rightarrow \infty} \prod_{i=1}^n
[e^{i\hat{H} T/n}e^{-ik f(t_i) \hat{A}T/n} ].
\end{eqnarray}
 The operator $\hat{U}_f$  is the generator of an
 one-parameter family of transformations
\begin{eqnarray}
 -i\frac{\partial}{\partial s} \hat{U}_f(s,k)  = [ \hat{H} + k f(s)
 \hat{A}] \hat{U}_f(s,k).
\end{eqnarray}
This implies that
\begin{eqnarray}
\hat{U}_f(T,k) = {\cal T} e^{i \int_0^T dt (H + k f(t) \hat{A})},
\label{Uf}
\end{eqnarray}
 where
${\cal T}$ signifies the  time-ordered expansion for the
exponential. The construction of $\hat{C}_U$ then is mathematically
identical to the determination of a propagator in presence of a
time-dependent external force proportional to $\hat{A}$.

For $f(t) = \frac{1}{T} \chi_{[0, T]}(t)$
 we  obtain
\begin{eqnarray}
\hat{C}_U =  \int \frac{dk}{2\pi} \,\tilde{\chi}_U(k) e^{i \hat{H} T
+ i k \hat{A}},
\end{eqnarray}
that has been constructed through path-integrals for specific
choices of the operator $\hat{A}$ in \cite{So98, LS00, BoHa05}.

If $f(t)$ has support in the interval $[t, t'] \subset [0, T]$ then
\begin{eqnarray}
\hat{C}_U = e^{-i \hat{H}t} \int_U da \; \left( \int \frac{dk}{2
\pi} e^{-i ka} {\cal T} e^{i \int_{t}^{t'} ds (H + k f(s)
\hat{A})}\right) e^{i\hat{H}(T-t')}.
\end{eqnarray}
We note that outside the interval $[t, t']$ only the Hamiltonian
evolution contributes to $\hat{C}_U$ outside the interval $[t, t']$.

It will be convenient to represent the proposition about the
time-averaged value of $\hat{A}$ by the operator
\begin{eqnarray}
\hat{D}(a) : =  e^{- i \hat{H}T} \hat{C}(a),
\end{eqnarray}

or else
\begin{eqnarray}
\hat{D}(a) = \int \frac{dk}{2 \pi} e^{ -i ka} \;{\cal T} e^{ik
\int_0^T dt f(t) \hat{A}(t)}, \label{toe}
\end{eqnarray}
where $\hat{A}(t)$ is the Heisenberg-picture operator $e^{i
\hat{H}t} \hat{A} e^{-i \hat{H}t}$.

If $[\hat{H}, \hat{A}] = 0$, then
\begin{eqnarray}
\hat{U}_f(T,k) = e^{i \hat{A}\int_0^T dt f(t) }.
\end{eqnarray}
 Hence,
 \begin{eqnarray}
\hat{D}_U := \int_U da \hat{D}(a) = \chi_U[\hat{A} \int_0^T dt
f(t)].
\end{eqnarray}
 When we use $f(t)$ to represent time-smearing, it
is convenient to require that $\int_0^T dt f(t)) = 1$ in order to
avoid any rescaling in the values of the observable. Then $\hat{D}_U
= \chi_U(\hat{A})$. We conclude therefore that the operator
representing time-averaged value of $\hat{A}$ coincides with the one
representing a single-time value of $\hat{A}$.

\paragraph{The limit of large coarse-graining.} If we integrate
$\hat{D}(a)$ over a relatively large  sample set $U$  the integral
over $dk$ is dominated by small values of $k$. To see this, we
approximate the integration over a subset of the real line of width
$\Delta$ centered around $a = a_0$, by an integral with a smeared
characteristic function $\exp[- (a-a_0)^2/2 \Delta^2]$. This leads
to
\begin{eqnarray}
\hat{D}_U = \sqrt{2\pi} \Delta \int \frac{dk}{2
 \pi} e^{- \Delta^2 k^2/2} \, {\cal T} e^{i \int_0^T
dt f(t) \hat{A}(t)}
\end{eqnarray}
 that is dominated by values of $k \sim
\Delta^{-1}$ .

The term $k f(t)$ in the time-ordered exponential of Eq. (\ref{toe})
is structurally similar to a coupling constant. Hence, for
sufficiently large values of $\Delta$ we  write
\begin{eqnarray}
{\cal T} e^{i \int_0^T dt f(t) \hat{A}(t)} \simeq e^{i \int_0^T dt
f(t) \hat{A}(t)},
\end{eqnarray}
 i.e., the zero-th loop order contribution to the
time-ordered exponential dominates. We therefore conclude that
\begin{eqnarray}
\hat{D}_U \simeq \chi_U \left[ \int_0^T dt f(t) \hat{A}(t) \right].
\label{111}
\end{eqnarray}
 $\hat{D}_U$ is almost equal to a spectral element of the
time-averaged Heisenberg-picture operator $\int_0^T dt f(t)
\hat{A}(t)$. This generalises the result of \cite{BoHa05}, which was
obtained for configuration space variables at the limit $\hbar
\rightarrow 0$.

We  estimate the leading order correction to the approximation
involved in Eq. (\ref{111}). The immediately larger contribution to
the time-ordered exponential of Eq. (\ref{toe}) is
\begin{eqnarray}
\frac{k^2}{2} \int_0^T ds \int_0^s ds' \, f(s) f(s') \, [\hat{A}(s),
\hat{A}(s')].
\end{eqnarray}
The contribution of this term must be much smaller than the first
term in the expansion of the time-ordered exponential's, namely $k
\int_0^T ds f(s) \hat{A}(s)$. We write the expectation values of
these operators on a vector $| \psi \rangle$ in order to obtain the
following condition
\begin{eqnarray}
|\int_0^T ds \int_0^s ds' f(s) f(s') \langle \psi|[\hat{A}(s),
\hat{A}(s')]| \psi \rangle| << \Delta \; | \int_0^T ds \langle \psi|
\hat{A}(s) | \psi \rangle|. \label{condit}
\end{eqnarray}

The above condition is satisfied rather trivially for bounded
operators if $||\hat{A}|| << \Delta$. In that case, the operator
$\hat{C}_U$ captures little, if anything, from the possible values
of $\hat{A}$. In the generic case however, Eq. (\ref{condit}) is to
be interpreted as a {\em condition} on the state $|\psi \rangle$.
Eq. (\ref{111}) provides a good approximation if the two-time
correlation functions of the system are relatively small.

Furthermore, if the function $f(t)$ corresponds to  weighted
averaging, i.e., if $f(t)  \geq 0 $,  and if $f$ does not have any
sharp peaks, then the condition $ \int_0^T dt f(t) = 1$ implies that
the values of $f(t)$ are of the order $\frac{1}{T}$.

We denote by $\tau$  the correlation time of $\hat{A}(s)$, i.e. the
values of $|s-s'|$ for which $|\langle \psi|[\hat{A}(s),
\hat{A}(s')]| \psi \rangle|$ is appreciably larger than zero. Then
at the limit
 $T >> \tau$  the
 left-hand side of Eq. (\ref{condit}) is of the order $O\left(
 \frac{\tau^2}{T^2}\right)$. Hence, for sufficiently large values of
 $T$
 one expects that Eq. (\ref{111}) will be satisfied with a fair degree of accuracy.

 The argument above does not hold if $f$ is allowed to take
 on negative values, which is the case for the velocity
 samplings that we  consider in section 4.

\subsection{Examples}
We study some interesting examples of  class operators corresponding
to time-smeared quantities. In particular, we  consider the
time-smeared position for a particle.and  a simple system that is
described by a finite-dimensional Hilbert space.

\subsubsection{Two-level system}
In a two-level system described by the Hamiltonian $\hat{H} = \omega
\hat{\sigma}_z$, we consider  time-averaged samplings of the values
of the operator $\hat{A} = \hat{\sigma}_x$. We compute
\begin{eqnarray}
\hat{U}(k,T) =  \cos \sqrt{k^2 + \omega^2 T^2}\, \hat{1} + i \,
\frac{\sin \sqrt{k^2 + \omega^2 T^2}}{\sqrt{k^2 + \omega^2 T^2}} ( k
\hat{\sigma_x} + \omega T \hat{\sigma}_z).
\end{eqnarray}

Then the class operator $\hat{C}(a)$ is
\begin{eqnarray}
\hat{C}(a) =  \frac{ \omega T}{2\sqrt{1 - a^2}} J_1( \omega T
\sqrt{1 - a^2})\, \hat{1} + \frac{a \omega T}{2 \sqrt{1-a^2}} J_1
(\omega T \sqrt{1 - a^2}) \hat{\sigma}_x  \nonumber \\
+ \frac{i \omega T}{2}  J_0(\omega T \sqrt{1 - a^2}) \hat{\sigma}_z,
\end{eqnarray}
where $J_n$ stands for the Bessel function of order $n$. Note that
the expression above holds for $|a| \leq 1$. For $|a|
> 1$, $\hat{C}(a) = 0$, as is expected by the fact that
$||\hat{\sigma}_x|| = 1$.

\subsubsection{Position samplings}
The case $\hat{A} = \hat{x}$ for ordinary time-averaging ($f(t) =
\frac{1}{T}$) has been studied in \cite{LS00, BoHa05} using path
integral techniques. Here we generalise these results by considering
the case of a general smearing function $f(t)$.

We consider the case of a harmonic oscillator of mass $m$ and
frequency $\omega$.  The determination of the propagator
$\hat{U}_f(T,k)$ for a harmonic oscillator  acted by an external
time-dependent force is well-known. It leads to the following
expression for the operator $\hat{D}(a)$
\begin{eqnarray}
\langle x|\hat{D}(a)|x' \rangle = \frac{m \omega}{2 \pi B_f \sin
\omega T} \exp \left[ \frac{ -i m \omega}{2 \sin \omega T} \left(
\cos \omega T (x'^2 - x^2) - 2 x x' \right)+  \right. \nonumber \\
 \left.
\left. \frac{2}{B_f} (A_f x' + a)(x'-x) - \frac{ 2 \omega C_f}{B_f^2
\sin \omega T} (x - x')^2 \right) \right],
\end{eqnarray}
where
\begin{eqnarray}
A_f :&=& \frac{1}{\sin \omega T}  \int_0^T ds \,  \sin \omega s \, f(s)\\
B_f :&=& \frac{1}{\sin \omega T} \int_0^T ds \, \sin \omega(T-s) f(s) \\
C_f :&=& \frac{1}{ \omega \sin \omega T} \int_0^T ds \, \sin \omega
(T-s) f(s) \int_0^s ds' \, \sin \omega s'\, f(s').
\end{eqnarray}
The corresponding operators for the free particle is obtained at the
limit $\omega \rightarrow 0$
\begin{eqnarray}
\langle x|\hat{D}(a)|x' \rangle = \frac{m}{2 \pi B_f T} \exp \left[
\frac{-im}{2T} \left( (x'^2 -x^2) + \frac{2}{B_f} (A_f x' - a)(x' -
x) - \frac{2C_f}{B_f^2 T} (x' - x)^2 \right) \right], \label{freep}
\end{eqnarray}
where
\begin{eqnarray}
A_f &=& \frac{1}{T}  \int_0^T ds \, s \, f(s)  \\
B_f &=& \frac{1}{ T} \int_0^T ds \, (T-s) f(s) \\
C_f &=& \frac{1}{ T} \int_0^T ds \,  (T-s) f(s) \int_0^s ds' \, s'
f(s').
\end{eqnarray}

\section{Probability assignment}

\subsection{The decoherence functional}

 For a pair of histories $(U,
 U')$ that correspond to different samplings of the time-smeared
 values of  $\hat{A}$ the decohrence functional $ d(U,U')$ is
 \begin{eqnarray}
d(U,U') = Tr \left(\hat{D}^{\dagger}_U e^{-i \hat{H}T} \hat{\rho}_0
e^{i \hat{H}T} \hat{D}_{U'} \right).
 \end{eqnarray}
From the expression above, we can read the probabilities that are
associated to any set of alternatives that satisfies the decoherence
condition. In section 2, we established  that in the limit of large
coarse-graining, or for very large values of time $T$, the operators
$\hat{D}_U$ approximate projection operators. Hence, if we partition
the real line of values of $A_f$ into sufficiently large exclusive
sets $U_i$ the decoherence condition will be satisfied. A
probability measure will be therefore defined as
\begin{eqnarray}
p(U_i) = Tr \left[\chi_{U_i}\left(\int_0^T dt f(t) \hat{A}(t)\right)
e^{-i \hat{H}T }\hat{\rho}_0 e^{i \hat{H}T} \right].
\end{eqnarray}
This is the same as in the case of a single-time measurement of the
observable $\int_0^T dt f(t) \hat{A}(t)$ taking place at time $t =
T$. For further discussion, see \cite{BoHa05}.
\subsection{Probabilities for measurement outcomes}

Next, we show that the class operators $\hat{C}(a)$ can be employed
in order to define a POVM for a measurement with finite duration.
For this purpose, we consider a simple measurement scheme.
 We
assume that the  system  interacts
 with a measurement device characterised by a continuous pointer
basis $|x \rangle$. For simplicity, we assume that the self-dynamics
of the measurement device is negligible. The interaction between the
measured system and the apparatus is described by  a Hamiltonian of
the form
\begin{eqnarray}
\hat{H}_{int} =  f(t)  \hat{A} \otimes \hat{K},
\end{eqnarray}
where $\hat{K}$ is the `conjugate momentum' of the pointer variable
$\hat{x}$
\begin{eqnarray}
 \hat{K} = \int dk \,k \,|k \rangle
\langle k|,
\end{eqnarray}

 where $\langle x| k \rangle = \frac{1}{\sqrt{2\pi}} e^{-ikx}$.
The initial state of the apparatus (at $t = 0$) is assumed to be
$|\Psi_0 \rangle$ and the initial state of the system corresponds to
a density matrix $\hat{\rho}_0$.

With the above assumptions, the reduced density matrix of the
apparatus at time $T$ is
\begin{eqnarray}
\hat{\rho}_{app}(T) = \int dk \int dk' \; Tr
\left(\hat{U}^{\dagger}_f(T,k) \hat{\rho}_0 \hat{U}_f(T,k') \right)
\langle k |\Psi_0 \rangle \langle \Psi_0|k' \rangle \; |k \rangle
\langle k'|,
\end{eqnarray}
where $\hat{U}_f(T,k)$ is given by Eq. (\ref{Uf}). Then, the
probability distribution over the pointer variable $x$ (after
reduction) is
\begin{eqnarray}
\langle x|\hat{\rho}_{app}(T)|x \rangle = \int \frac{dk dk'}{2 \pi}
 e^{-i(k-k')x} \langle k |\Psi_0 \rangle \langle \Psi_0|k' \rangle
\; Tr \left(\hat{U}^{\dagger}_f(T,k) \hat{\rho}_0 \hat{U}_f(T,k')
\right).
\end{eqnarray}

The probability that the pointer variable takes values within a set
$U$ is
\begin{eqnarray}
p(U) = tr \left(e^{-i \hat{H}T}\hat{\rho}_0 e^{i
\hat{H}T}\hat{\Pi}_U \right),
\end{eqnarray}
where

\begin{eqnarray}
\hat{\Pi}_U  = \int_U dx \; \hat{D}(w^*_x) \hat{D}^{\dagger}(w_x) :=
\int_U dx \, \hat{\Pi}_x, \label{central}
\end{eqnarray}

where  $ w_x(a):= \langle x-a|\Psi_0 \rangle$ and where we employed
the notation

\begin{eqnarray}
\hat{D}(w_x) = \int da \, w_x(a) \hat{D}(a),
\end{eqnarray}

The operators $\hat{\Pi}_U$ define a POVM for the time-extended
measurement of $\hat{A}$: they are positive by construction, they
satisfy the property $ \hat{\Pi}_{U_1 \cup U_2} = \hat{\Pi}_{U_1} +
\hat{\Pi}_{U_2}$, for $U_1 \cap U_2 = \emptyset$ and they are
normalised to unity
\begin{eqnarray}
\hat{\Pi}_{\bf R} = \int_{\bf R} dx \, \hat{\Pi}_x = 1.
\end{eqnarray}

Note that the smearing of the class-operators is due to the spread
of the wave function of the pointer variable.

In what follows we employ for convenience a Gaussian  function
\begin{eqnarray}
w(a) = \frac{1}{(2 \pi \delta^2)^{1/4}} e^{- \frac{a^2}{4
\delta^2}}. \label{ww}
\end{eqnarray}
In the  free-particle case, the class operators in Eq. (\ref{freep})
lead to the following POVM
\begin{eqnarray}
\langle y|\hat{\Pi}_x|y' \rangle = \frac{m}{\sqrt{2} \pi A_f T} \exp
\left[ -\left(\frac{m^2 \delta^2}{2 A_f^2 T^2} + \frac{A_f^2}{8
\delta^2}(1 - \frac{2C_f}{A_f^2T})^2\right) (y - y')^2 +
\frac{im}{A_fT} x(y' - y) \right]. \label{fff}
\end{eqnarray}

In Eq. (\ref{fff}), we  chose an even time-averaging function, i.e.
$f(s) = f(T-s)$, in which case $A_f = B_f$.

The POVM in Eq. (\ref{central}) may also be constructed without
reference to a specific model for the measurement device. In
particular, we partition the space of  values for $A_f$ into sets of
width $\delta$ and employ the expression Eq. (\ref{pmeas}) for the
ensuing probabilities.  It is easy to show that these probabilities
are reproduced---up to terms of order $O(\delta)$---by a POVM of the
form Eq. (\ref{central}), with the
 smearing function $w$ of Eq. (\ref{ww})
\footnote{ The proof follows closely an analogous one in
\cite{Ana05}).}.

If we restrict our considerations to the above measurement model,
then there is no way we can interpret the POVM of Eq.
(\ref{central}) as corresponding to values of $A_f$,  This
interpretation is  possible by the explicit construction  and by the
identification (see Sec. 2) of the class operators $\hat{C}(a)$ as
the only mathematical objects that correspond to such time-averaged
alternatives.

\section{More general samplings}

\subsection{Velocity Vs momentum}
Within the context of the History Projection Operator scheme,
Savvidou showed that histories of momentum differ in general from
histories of velocity, in the sense that they are represented by
different mathematical objects \cite{Sav99}. The corresponding
probabilities are also expected to be different. In single-time
quantum theory the notion of velocity (that involves differentiation
with respect to time) cannot be distinguished from the notion of
momentum. However, when we deal with histories, time differentiation
is defined {\em independently} of the evolution laws. One may
therefore consider alternatives corresponding to different values of
velocity.

In particular, if $x_f  = \int_0^T dt x_t f(t)$ denotes the
time-smeared value of the position variable, we define the
time-smeared value of the corresponding velocity variable as
\begin{eqnarray}
\dot{x}_f  := - x_{\dot{f}},
\end{eqnarray}
 provided that the function $f$
satisfies $f(0) = f(T) = 0$.

Notice here that when we measure the time-averaged value of an
observable within a time-interval $[0, T]$, we employ positive
functions $f(t)$ that are $\cap$-shaped and that they satisfy
$\int_0^Tdt f(t) = 1$. Such functions correspond to the intuitive
notions of averaging the value of a quantity with a specific weight.

 However,  to determine the time-average  velocity---weighted by a
positive and normalised function $f$---one has to smear the
corresponding position variable with the function $\dot{f}(t)$ that
in the general case  is neither positive nor normalised. Therefore
{\em the form of the smearing function determines the physical
interpretation of the observable we consider} \cite{Sav05}.

Next, we compare the class operators corresponding to the average of
velocity and of momentum, with a common weight $f$. We denote the
velocity class operator as

\begin{eqnarray}
\hat{D}^{\dot{x}}(a) = \int \frac{dk}{2 \pi} e^{-ika} {\cal T} e^{i
\int_0^T dt \,\dot{f}(t) \hat{x}(t)},
\end{eqnarray}
and the momentum class operators as
\begin{eqnarray}
 \hat{D}^{p}(a)  = \int
\frac{dk}{2 \pi} e^{-ika} {\cal T} e^{i \int_0^T dt \, f(t)
\hat{p}(t)}.
\end{eqnarray}

At the limit of large coarse-graining, the operator
$\hat{D}^{\dot{x}}_U :=\int_U da \, \hat{D}^{\dot{x}}(a)$ is
approximately equal to
\begin{eqnarray}
\hat{D}^{\dot{x}}_U = \chi_U(\int_0^T dt \dot{f}(t) \hat{x}(t)) =
\chi_U( \frac{1}{m} \int_0^T dt \, f(t) \hat{p}(t)),
\end{eqnarray}
 {\em i.e.},  the class-operator for
time-averaged momentum coincides with that for time-averaged
velocity. This result reproduces the classical notion that $p = m
\dot{x}$. However,  the limit of large coarse-graining may be
completely trivial if the temporal correlations of position are
large.

For the case of a free particle, with the convenient choice  $f(t) =
\frac{\pi}{T}  \sin\frac{\pi t}{T}$, we obtain
\begin{eqnarray}
\hat{D}^p_U &=& \int_U dp \, |p \rangle \langle p|, \\
\hat{D}^{\dot{x}}_U &=& \int_U da \; \left(\sqrt{\frac{4 i m
T}{\pi^3}} \int dp \;  e^{i\frac{4 m T}{\pi^2} (a - p/m)^2} \, |p
\rangle \langle p| \right). \label{ddx}
\end{eqnarray}

It is clear that the alternatives of time-averaged momentum are
distinct from those of time-averaged velocity. Still, at the limit
$T \rightarrow \infty$, $ \hat{D}^p_U =
 m \hat{D}^{\dot{x}}_U $.

The  POVM corresponding to Eq. (\ref{ddx}) is
\begin{eqnarray}
\hat{\Pi}^{\dot{x}}(v) = \frac{1}{\sqrt{2 \pi \sigma^2(T)}}
 \int dp \; \exp
\left[ - \frac{1}{2 \sigma_v^2(T)} (v - p/m)^2 \right]\, |p \rangle
\langle p|, \label{POVMvelocity}
\end{eqnarray}
where $\sigma_v^2(T) = \delta^2 + \frac{\pi^4}{2^{8} m^2 T^2
\delta^2}$.

The POVM of Eq. (\ref{POVMvelocity}) commutes with the momentum
operator. One could therefore claim that it corresponds to  an
unsharp measurement of momentum. However, the commutativity of this
POVM with momentum follows only from the special symmetry of the
Hamiltonian for a free-particle, it does not hold in general.
Moreover, at the limit of  small $T$, the distribution corresponding
to Eq. (\ref{POVMvelocity}) has a very large mean deviation. Hence,
even for a wave-packet narrowly concentrated in momentum, the spread
in measured values is  large. Note that at the limit $T \rightarrow
0$, the deviation $\sigma^2_v(T) \rightarrow \infty$ and the POVM
(\ref{POVMvelocity}) tends weakly to zero. For $T
>> (m \delta^2)^{-1}$, then $\sigma_v^2(T) \simeq \delta^2$ and the
velocity POVM is identical to one obtained by an instantaneous
momentum measurement.

The results of section 3.2 suggest the different measurement schemes
that are needed for the distinction of velocity and momentum. For a
momentum measurement the interaction Hamiltonian should be of the
form
\begin{eqnarray}
\hat{H}_{int}^{p} =  f(t)  \, \hat{p}\, \otimes \hat{K},
\end{eqnarray}
where $f(t)$ is a $\cap$-shaped positive-valued function. For a
velocity measurement the  interaction Hamiltonian is
\begin{eqnarray}
\hat{H}_{int}^{\dot{x}} =  - \dot{f}(t)  \, \hat{x} \otimes \hat{K}.
\end{eqnarray}

The two Hamiltonians differ not only on the coupling but also on the
shape of the corresponding smearing functions: $\dot{f}(t)$ takes
both positive and negative values and by definition it satisfies
$\int_0^T \dot{f}(t) = 0$. The description above suggests that
momentum measurements can be obtained by coupling a charged particle
to a magnetic field pulse, while velocity measurements can be
obtained by a coupling to an electric field pulse of a different
shape. The possibility of designing realistic experiments that could
distinguish between the momentum and the velocity content of a
quantum state will be discussed elsewhere.

\subsection{Lagrangian action}

One may also consider samplings corresponding to the values of the
Lagrangian action of the system $\int_0^T dt L(x, \dot{x})$, where
$L$ is the Lagrangian. In this case the results can be easily
expressed in terms of Feynman path integrals: it is straightforward
to demonstrate---see Ref. \cite{So98}---that these coincide with
samplings of the Hamiltonian, and that the corresponding POVM is
that of energy measurements.

\subsection{Phase space properties}

It is possible to construct class-operators (and corresponding
POVMs) for more general alternatives that involve phase-space
variables. To see this, we consider a set of coherent states $|z
\rangle$ on the Hilbert space, where $z$ denotes points of the
corresponding classical phase space. The finest-grained histories
corresponding to an $n$-time coherent state path $z_0, t_0, z_1,
t_1, \ldots z_n, t_n$, with $t_i - t_{i-1} = \delta t$ are
represented by the class operator
\begin{eqnarray}
\hat{C}_{z_0, t_0; z_1, t_1; \ldots; z_n, t_n} = | z_0 \rangle
\langle z_0 |e^{i \hat{H} \delta t}|z_1 \rangle \langle z_1|
e^{i\hat{H} \delta t} |z_2 \rangle \cdots \langle z_{n-1}|e^{i
\hat{H} \delta t}|z_n \rangle \langle z_n|.
\end{eqnarray}
We use the standard Gaussian coherent states, which are defined
through an inner product
\begin{eqnarray}
\langle z|z' \rangle = e^{ - \frac{|z|^2}{2} - \frac{|z'|^2}{2} +
z^* z'}.
\end{eqnarray}
Then,  at the limit of  small $\delta t$
\begin{eqnarray}
\hat{C}_{z_1, t_1; z_2, t_2; \ldots; z_n, t_n}  = |z_0 \rangle
\langle z_n |\; \exp \left(\frac{|z_n|^2}{2} - \frac{|z_0|^2}{2} -
\sum_{i=1}^n z^*_i(z_i - z_{i-1}) + i \delta t \,h(z^*_i, z_{i-1})
\right),
\end{eqnarray}
where $ h(z^*, z) = \langle z|\hat{H}|z  \rangle$.  Following the
same steps as in section 2.1 we construct the class operator
corresponding to different values of an observable $A(z_0, z_1,
\ldots, z_n)$. If the observable is ultra-local, i.e., if it can be
written in the form $\sum_i f(t_i) a(z_i)$, then the results reduce
to those of section 2.1 for the time-smeared alternatives of an
operator.

However, the function in question may involve  time derivatives of
phase space variables (at the continuous limit), in which case it
will be rather different from  the ones we considered previously.
For a generic function $F(z_i)$ we obtain the following class
operator
 that corresponds to  the value $F =a$
\begin{eqnarray}
\langle z_0| \hat{C}(a)|z_f \rangle  = \int \frac{dk}{2 \pi} e^{-i
ka} \; \lim_{n \rightarrow \infty} \int  [dz_1]
\ldots [dz_{n-1}] \hspace{3cm} \nonumber \\
\times \exp \left[\frac{|z_n|^2}{2} - \frac{|z_0|^2}{2} + \sum_i
z_i(z^*_i - z^*_{i-1}) + i \delta t \,h(z^*_i, z_{i-1})  - i k
F[z_i]\right]. \label{cc}
\end{eqnarray}
The integrations over $[dz_i]$ defines  a coherent-state
path-integral at the continuous limit. However, if $F[z_i]$ is not
an ultra-local function, the path integral does not correspond to a
unitary operator of the form ${\cal T} e^{\int_0^T dt \hat{K}_t}$,
for some family of self adjoint operators $\hat{K}_t$. In this
sense, the consideration of phase space paths provides alternatives
that do not reduce to those studied in Section 2. Note however, that
these alternatives cannot be defined in terms of projection
operators; nonetheless the corresponding class operators can be
employed to define a POVM using Eq. (\ref{central}).

The simplest non-trivial example of a non-ultralocal function is the
Liouville term of the phase space action ( for its physical
interpretation in the histories theory see \cite{Sav99})
\begin{eqnarray}
 V :=i  \int_0^T dt \dot{z}^* z.
\end{eqnarray}
It is convenient to employ the discretised expression $V =
 i \sum_{i=1}^n z_i(z^*_i -
z^*_{i-1})$. Its substitution in Eq. (\ref{cc}) effects a
multiplication of the Liouville term in the exponential by a factor
of $1+k$.

For an harmonic oscillator Hamiltonian $h(z^*,z) = \omega z^* z$,
and the path integral can be explicitly computed yielding the
unitary operator $e^{\frac{i}{1+k} \hat{H}T}$. Hence,
\begin{eqnarray}
\hat{C}(a) = \int \frac{dk}{2 \pi} e^{-i ka} e^{\frac{i}{1+k}
\hat{H}T} = s_a(\hat{H}),
\end{eqnarray}

where $s_a(x): = \int \frac{dk}{2 \pi} e^{-ika + i \frac{x}{1+k}}$.
The class-operator $\hat{C}(a)$ corresponding to the values of the
function $V$ is then a function of the Hamiltonian.

\section{Conclusions}

We studied the probability assignment for time-extended
measurements. We constructed of the class operators $\hat{C}(a)$,
which  correspond to time-extended alternatives for a quantum
system. We showed that these operators can be employed to construct
POVMs describing the probabilities for time-averaged values of a
physical quantity. In light of these results, quantum mechanics has
room for measurement schemes that distinguish between momentum and
velocity. Finally, we demonstrated that a large class of
time-extended phase space observables may be explicitly constructed.

\section*{Acknowledgements}
C.A. was funded by a Pythagoras II grant (EPEAEK). N.S. acknowledges
support from the EP/C517687 EPSRC grant.

\end{document}